\documentstyle[11pt]{amsart}

\newcommand{\FL}{\bf L}
\newcommand{\half}[1]{\lfloor#1/2\rfloor}
\newcommand{\sbz}{\Sigma_0^b}
\newcommand{\sbi}[1]{\Sigma_{#1}^b}
\newcommand{\pbi}[1]{\Pi_{#1}^b}
\newcommand{\dbi}[1]{\Delta_{#1}^b}

\newcommand{\Bit}{{\rm Bit}}
\newcommand{\all}{\forall}
\newcommand{\ext}{\exists}
\newcommand{\acca}{{\rm AC}^0{\rm CA}}
\newcommand{\logi}[1]{\log^{(#1)}}

\newcommand{\rest}{\lceil}
\newcommand{\seq}[1]{\langle#1\rangle}

\newtheorem{definition}{Definition}
\newtheorem{theorem}{Theorem}
\newtheorem{lemma}{Lemma}
\newtheorem{corollary}{Corollary}
\newtheorem{proposition}{Proposition}
\newtheorem{problem}{Problem}
\newtheorem{remark}{remark}

\begin{document}

\title[Weak Length Induction]{Weak Length Induction and Slow Growing Depth Boolean Circuits}
\author{Satoru Kuroda\\
Toyota National College of Technology}
\address{2-1, Eisei-cho, Toyota, 471-8525, Japan}
\email{satoru@@math.human.nagoya-u.ac.jp}
\date{}

\maketitle

\vspace{2cm}
\noindent

\begin{abstract}
We define a hierarchy of circuit complexity classes $LD^i$, whose depth 
are the inverse of a function in Ackermann hierarchy. Then we introduce 
extremely weak 
versions of length induction called $L^{(m)}IND$ and construct a bounded 
arithmetic theory $L^i_2$ whose provably total functions exactly corresponds 
to functions computable by $LD^i$ circuits. Finally, we prove a 
non-conservation result between $L^i_2$ and a weaker theory $AC^0CA$ which 
corresponds to $AC^0$. Our proof utilizes KPT witnessing theorem.
\end{abstract}

\section{Introduction}
A fundamental problem of boolean circuit complexity is whether the NC/AC 
hierarchy collapses. And the only known separation is that $AC^0\neq NC^1$. 
In words, there is a gap between constant depth and logarithmic depth. 
This separation was done by Furst Saxe and Sipser \cite{fss} and H\aa stad 
\cite{hastad} who showed that the parity function is not in $AC^0$ (see also 
Ajtai \cite{ajtai}). Although 
there are some $NC^1$ functions which are not in $AC^0$, not much is known 
about the gap between these two classes.

In this note we define a hierarchy which lies between $AC^0$ and $AC^1$. 
Let $\logi{m}x=\log(\log(\cdots\log x))$ ($m$ times) and define the class 
$LD^i$ to be those which are computable by polynomial size 
$\left(\logi{i} n\right)^{O(1)}$ depth circuits. Then 
$$
AC^0\subseteq\cdots\subseteq LD^3\subseteq LD^2\subseteq AC^1\subseteq LD^1
$$
holds. Natural questions to ask about the LD hierarchy are whether 
$LD^i=LD^j$ for some $i\neq j$, and even whether $AC^0=LD^i$ or $LD^i=AC^1$ 
for some $i$. Furthermore, there lies several other interesting classes 
such as ALOGTIME, LOGSPACE or NLOGSPACE along this hierarchy, and inclusion 
between these classes and $LD^i$ also seems to be open.

In order to attack these problems we shall give a proof theoretical 
counterpart to the LD hierarchy. In \cite{buss} S.~Buss defined weak system of 
arithmetic whose provably total functions correspond to the polynomial 
hierarchy. His theory $S^i_2$ utilizes weak form of induction scheme 
which is called length induction (LIND). By modifying the length induction 
scheme in much weaker form, we can characterize theories which correspond 
to the LD hierarchy. 

The depth-bounding function $\log^{(m)}n$ is known to be an inverse of 
a function in the Ackermann hierarchy,(cf. \cite{bb}) so it is a very slow 
growing function. 
Nevertheless, in the last section we show that their proof-theoretical 
counterparts are distinct. Namely, we shall use KPT witnessing theorem 
to separate our theory $L^i_2$ from the theory $\acca$ whose definable 
functions are those in $AC^0$. Although our result cannot be used to 
separate the corresponding complexity classes $LD^i$ and $AC^0$, it is 
strong enough an evidence to conjecture that $AC^0\neq LD^i$. 

\section{Basic Definitions}

Throughout this note we assume that all logarithms are base 2. Define the 
function $\log^{(m)}x$ for $m\in\omega$ and $|x|_m$ by
$$
\begin{array}{rl}
  \log^{(1)}x&=\log x,\\
  \log^{(m+1)}x&=\log\left(\log^{(m)}x\right),
\end{array}
$$
and 
$$
\begin{array}{rl}
  |x|_1&=\lceil\log (x+1)\rceil,\\
  |x|_{m+1}&=\left||x|_m\right|.
\end{array}
$$
Also we denote $|x|_1$ by $|x|$ which is equal to the length of binary 
representation of $x$.

We treat functions and sets of both natural 
numbers and binary strings. Numbers are often identified with binary strings 
by considering their binary expansions and conversely, binary strings are 
identified with corresponding natural numbers. The set of binary strings 
is denoted by $\{0,1\}^*$ and binary strings with length $n$ by $\{0,1\}^n$. 
For a natural number $x$ let $|x|$ be its length in binary. For any 
complexity class $C$ we mean a class of functions and sets (predicates) 
are identified with their characteristic functions. 

As usual, a circuit is a directed acyclic graph with each node labeled by 
either $x_1,x_2,\ldots,x_n$, $\land$, $\lor$, $\neg$. Internal nodes are 
called gates and labeled by either $\land$, $\lor$, $\neg$. Nodes without 
input edges are called input and labeled by one of $x_1,x_2,\ldots,x_n$. 
The size of a circuit is the number of gates and the depth is the length of 
the longest path in it. The fan-in of a gate is the number of input edges 
and the fan-in of the circuit is the maximum of fan-in of gates in it. 

We assume that every circuit has only one output so that it computes a 
predicate. We say that a circuit family $C_1,\ldots,C_m$ computes a function 
$f:\{0,1\}^n\rightarrow\{0,1\}^m$ if its bitgraph is computed by each circuit 
$C_i$. Or equivalently, putting all $C_i$'s altogether yields a multi-output 
circuit that computes $f$. Hence we can assume that any finite 
function $f:\{0,1\}^n\rightarrow\{0,1\}^m$ is computed by a single circuit. 

Furthermore, we assume that all circuit families are $U_{E^*}$-uniform.

\begin{definition}
  A function $f:\{0,1\}^+\rightarrow\{0,1\}^+$ is computed by a circuit 
family $\{C_n\}_{n\in\omega}$ if for all $n\in\omega$, $f\rest_n$($f$ 
restricted to the set $\{0,1\}^n$) is computed by $C_n$. 
\end{definition}

First we give a precise definition of the class $LD^i$. 
\begin{definition}
  $LD^i$ is the class of functions computed by some $n^{O(1)}$ size, 
$\left(\logi{i} n\right)^{O(1)}$ depth circuits of unbounded fan-in. 
\end{definition}
\begin{remark}
$LD^1=NC$.  
\end{remark}

Let $AC^i$ be the class of functions computable by some polynomial size, 
$O((\log n)^i)$ depth circuits. Then 
\begin{proposition}
For $i\geq 2$, $AC^0\subseteq LD^i\subseteq AC^1$.
\end{proposition}
\begin{remark}
Since $AC^0\neq AC^1$, at least either one of $AC^0\neq LD^i$ or 
$LD^i\neq AC^1$ holds for any $i\in\omega$. 
\end{remark}

We define a function algebra for $LD^i$ as follows: 
\begin{definition}
  INITIAL is the finite set of functions which consists of: $Z(x)=0$, 
$P^n_k(x_1,\ldots,x_n)=x_k$, $s_0(x)=2x$, $s_1(x)=2x+1$, 
$|x|=\lceil\log(x+1)\rceil$, $Bit(x,i)=\lfloor x/2^i\rfloor\mod 2$, 
$x\#y=2^{|x|\cdots|y|}$. 
\end{definition}
\begin{definition}
  \begin{enumerate}
  \item  A function $f$ is defined by Concatenation Recursion on Notation 
(CRN) from $g$, $h_0$ and $h_1$ if 
    $$
    \begin{array}{rl}
      f(0,\vec{y})&=g(\vec{y}),\\
      f(s_0(x),\vec{y})&=s_{h_0(x,\vec{y})}(f(x,\vec{y})),\mbox{ if }x\neq 0\\
      f(s_1(x),\vec{y})&=s_{h_1(x,\vec{y})}(f(x,\vec{y}))
    \end{array}
    $$
    provided that $h_i(x,\vec{y})\leq 1$ for all $x$ and $\vec{y}$ and 
    $i=0,1$. 
  \item Let $i\in\omega$. A function $f$ is defined by i-Weak Bounded 
    Recursion on Notation ($W^iBRN$) from $g$, $h_0$, $h_1$ and $k$ if
    $$
    \begin{array}{rl}
      F(0,\vec{y})&=g(\vec{y}),\\
      F(s_0(x),\vec{y})&=h_0(x,\vec{y},F(x,\vec{y})),\mbox{ if }x\neq 0\\
      F(s_1(x),\vec{y})&=h_1(x,\vec{y},F(x,\vec{y}))\\
      f(x,\vec{y})&=F(|x|_i,\vec{y}),
    \end{array}
    $$
    provided that $F(x,\vec{y})\leq k(x,\vec{y})$ for all $x,\vec{y}$. 
  \end{enumerate}
\end{definition}
In \cite{kuroda1} it is proved that 
\begin{definition}
  $LD^i$ is the smallest class of functions which contains INITIAL and 
  closed under composition CRN and $W^{i}BRN$ operations.
\end{definition}
Clote and Takeuti \cite{ct} showed that 
\begin{theorem}\label{thm:NC}
  $AC^0$ is the class of functions containing INITIAL and closed under 
composition and CRN operations.
\end{theorem}

We may use a weak successor-type recursion within $LD^i$.
\begin{proposition}
  $LD^i$ is closed under the following recursion operation:
$$
\begin{array}{rl}
F(0,\vec{y})&=g(\vec{y}),\\
F(x+1,\vec{y})&=h(x,\vec{y},F(x,\vec{y}),\\
f(x,\vec{y})&=F(|x|_{i+1},\vec{y}),
\end{array}
$$
where $F(x,\vec{y})\leq k(x,\vec{y})$.
\end{proposition}

Since it is unknown whether the majority function belongs to any of $LD^i$ 
for $i\geq 1$, we can extend the class $LD^i$ by adding majority gates as 
in the definition of $TC^0$. 
\begin{definition}
  For $i\geq 1$, $MD^i$ is defined as $LD^i$ but with additional majority 
gates.
\end{definition}

Then the characterization of $LD^i$ are modified as follows:
\begin{theorem}
  For $i\geq 1$, $MD^i$ is the smallest class of functions containing 
INITIAL and multiplication and closed under composition, CRN and $W^{i+1}BRN$ 
operations.
\end{theorem}

\section{The Theory $L^i_2$}
We assume that readers are familiar with basic notions of bounded arithmetic.
The first order language $\FL_1$ consists of function symbols 
$$
\begin{array}{l}
Z(x)=0, P^n_k(x_1,\ldots,x_n)=x_k, s_0(x)=2x, s_1(x)=2x+1\\
\half{x}, MSP(x,i)=\lfloor x/2^i\rfloor, |x|=\lceil\log_2(x+1)\rceil, 
x\#y=2^{|x|\cdot|y|}
\end{array}
$$
and a predicate symbol $\leq$. 

A quantifier is called {\it bounded} if it is either of the form 
$\all x\leq t$ or $\ext x\leq t$ and {\it sharply bounded} if it is either 
of the form $\all x\leq |t|$ or $\ext x\leq |t|$. A formula is bounded if 
all quantifiers are bounded and sharply bounded if all quantifiers are 
sharply bounded. 
$\sbz$ is the set of sharply bounded formulae. $\sbi{1}$ is the set of 
formulae in which all non-sharply bounded quantifiers are positive 
appearances of existential quantifiers and negative appearances of 
universal ones. $\pbi{1}$ is defined in the same way 
by replacing existential with universal. $\sbi{i}$ and $\pbi{i}$ ($i\geq 2$) 
are defined in an analogous manner.

Now let us state axioms that consist our theory $L^i_2$. 

BASIC is a finite set of axioms which define symbols in $\FL_1$. Let $\Phi$ 
be a set of formulae.
\begin{itemize}
\item $\Phi$-Bit-Comprehension:
$$
\ext y<2^{|t|}\all i<|t|\left[\Bit(i,x)=1\leftrightarrow\varphi(i)\right],
$$
\item $\Phi$-LIND:
$$
\varphi(0)\land\all x(\varphi(x)\rightarrow\varphi(x+1))\rightarrow
\all x\varphi(|x|),
$$
\item $\Phi$-$L^iIND$ (for $i\geq 2$):
$$
\varphi(0)\land\all x(\varphi(x)\rightarrow\varphi(x+1))\rightarrow
\all x\varphi(|x|_i),
$$
\item $\Phi$-$L^iMAX$:
$$
\ext x<a\varphi(x)\rightarrow
\ext x<a\left(\varphi(x)\land\all y<a(|x|_i<|y|_i\rightarrow\neg\varphi(y))
\right),
$$
\end{itemize}
where $\varphi\in\Phi$.

Our characterization is similar to that of Clote and Takeuti \cite{ct} 
who introduced the notion of {\it essentially sharply boundedness}. 
\begin{definition}
  Let $T$ be a theory. A formula $\varphi$ is essentially sharply bounded 
(esb) in $T$ if it belongs to the smallest class $C$ satisfying the 
following conditions:
\begin{itemize}
\item every atomic formula is in $C$. 
\item $C$ is closed under boolean connectives and sharply bounded 
quantifications.
\item If $\varphi_0,\varphi_1\in C$ and 
$$
\begin{array}{l}
T\vdash\ext x\leq s(\vec{a})\varphi_0(\vec{a},x)\\
T\vdash\all x,y\leq s(\vec{a})(\varphi_0(\vec{a},x)
                       \land\varphi_0(\vec{a},y)\rightarrow x=y)
\end{array}
$$ 
then 
$\ext x\leq s(\vec{a})(\varphi_0(\vec{a},x)\land\varphi_1(\vec{a},x))$ and 
$\all x\leq s(\vec{a})(\varphi_0(\vec{a},x)\rightarrow\varphi_1(\vec{a},x))$ 
are in $C$. 
\end{itemize}
A formula is $ep\sbi{1}$ in $T$ if it is of the form 
$\ext x_1\leq t_1\cdots\ext x_k\leq t_k\varphi(x_1,\ldots,x_k)$ where 
$\varphi$ is esb in $T$. 
\end{definition}

\begin{definition}
  $L^i_2$ is the $\FL_1$ theory which consists of the following axioms:
  \begin{itemize}
  \item BASIC
  \item Bit-Comprehension for esb formulae with respect to $L^i_2$
  \item esb-LIND for esb formulae with respect to $L^i_2$
  \item $ep\sbi{1}$-$L^{i+1}IND$.
  \end{itemize}
\end{definition}
The following fact will be used later.
\begin{proposition}
  $ep\sbi{1}$-$L^{i+1}MAX$ principle is a theorem of $L^i_2$.
\end{proposition}
\begin{pf}
  It is a well-known fact that $\sbi{i}$-LIND and $\pbi{i}$-LIND are 
equivalent over BASIC$+$open-LIND and its proof is directly applied 
to show that
$$
L^i_2\vdash 
ep\sbi{1}\mbox{-}L^{i+1}IND\leftrightarrow ep\pbi{1}\mbox{-}L^{i+1}IND,
$$
where $ep\pbi{1}$ is the set of formulae of the form 
$$
\all x_1\leq t_1\cdots\all x_k\leq t_k\varphi(x_1,\ldots,x_k)
$$
for esb formula $\varphi$ in $L^i_2$. 
It is also easy to see in $L^i_2$ that $ep\pbi{1}$-$L^{i+1}IND$ implies 
$ep\sbi{1}$-$L^{i+1}MAX$. 
\end{pf}
In \cite{kuroda1} the author defined a theory for $AC^0$ functions. 
\begin{definition}
The language $\FL_{AC^0}$ consists of function symbols for all $AC^0$ 
functions. $AC^0CA$ is the $\FL_{AC^0}$ theory which consists of the 
following axioms:
\begin{itemize}
\item defining axioms for each $f\in AC^0$,
\item $\sbz$-LIND.
\end{itemize}
\end{definition}

\begin{theorem}
  $AC^0CA$ is an universal theory.
\end{theorem}

It is also seen that $AC^0CA$ can be regarded as a subtheory of $L^i_2$ 
since
\begin{proposition}
  Let $f$ be a function symbol for an $AC^0$ function and let $L^i_2(f)$ 
be the theory $L^i_2$ extended by the symbol $f$ together with its defining 
axioms. Then $L^i_2(f)$ is an conservative extension of $L^i_2$. 
\end{proposition}

Hence in the following we assume that $L^i_2$ has all $AC^0$ functions 
in its language.

\begin{definition}
  A function $f$ is esb definable in a theory $T$ if there exists 
an esb formula $\varphi$ in $T$ such that 
$$
\begin{array}{l}
T\vdash\all x\ext y\varphi(x,y),\\
T\vdash\all x,y,z\varphi(x,z)\land\varphi(y,z)\rightarrow x=y,\\
\Bbb n\vdash\all x\varphi(x,f(x)).
\end{array}
$$
\end{definition}

The following immediately holds by the definition.
\begin{proposition}
  Let $\varphi(\vec{x},y)$ esb-define a function $f$ in a theory $T$. 
Then the following formulae are equivalent in $T(f)$
\begin{itemize}
\item $\ext x\leq s(\vec{a})(\varphi(\vec{a},x)\land\psi(\vec{a},x))$
\item $\all x\leq s(\vec{a})(\varphi(\vec{a},x)\rightarrow\psi(\vec{a},x))$
\item $\psi(\vec{a},f(\vec{a}))$. 
\end{itemize}
\end{proposition}
Hence esb formulae are $\dbi{1}$ with respect to the theory in concern 
and sharply bounded in the extended language.

\section{Definability of $LD^i$ Functions in $L^i_2$}
First we shall show that $LD^i$ functions can be defined in $L^i_2$. 

\begin{theorem}
  If $f\in LD^i$ then $f$ is $ep\sbi{1}$ definable in $L^i_2$. 
\end{theorem}
\begin{pf} 
The proof is by induction on the complexity of $f\in LD^i$. 

By the assumption that $L^i_2$ contains $AC^0CA$, all INITIAL functions 
are $ep\sbi{1}$ definable (actually esb definable) in $AC^0CA$ ,hence also 
in $L^i_2$. The same argument implies that the closure under composition 
and CRN are also proved within $AC^0CA$. 
So it suffices to show that $\sbi{1}$ definable functions of $L^i_2$ are 
closed under $W^{i}BRN$ operation. 

Let $f$ be defined by $W^{i}BRN$ from $g$, $h_0$, $h_1$ and $k$ each has 
$\sbi{1}$ definition in $L^i_2$. Let $\Phi(w)$ be the formula 
expressing that ``$w$ is a sequence of the computation of $f$''. Then it is 
readily seen that $\Phi\in ep\sbi{1}$ and 
$$
L^i_2\vdash\Phi(0)\land
\all x(\Phi(|w|)\rightarrow\Phi(|w|+1)).
$$
So by $\sbi{1}$-$L^{i+1}IND$ we have 
$L^i_2\vdash\all x\Phi(|w|_{i+1})$. 
Hence the $\sbi{1}$ formula $\Phi$ defines $f$ provably in $L^i_2$. 
\end{pf}

\section{Witnessing Proofs in $L^i_2$}
Now we shall show the converse to the previous theorem. Namely, All 
esb consequences of $L^i_2$ are witnessed by some $LD^i$ functions.
\begin{theorem}\label{thm:wit1}
Let $\varphi\in ep\sbi{1}$ be such that $L^i_2\vdash\all x\ext y\varphi(x,y)$. 
Then there exists a function $f\in LD^i$ such that 
$L^i_2\vdash\all x\varphi(x,f(x))$.
\end{theorem}

The proof is by the witnessing method. 

\begin{definition}
  Let $\varphi$ be an esb formula in $T$. Then we denote the equivalent 
sharply bounded formula (in the extended language) by $\varphi^{sb}$ (called 
sb version of $\varphi$). 
If $\varphi$ is $ep\sbi{1}$ of the form 
$\ext x_1\leq t_1\cdots\ext x_k\leq t_k\varphi(x_1,\ldots,x_k)$ where 
$\varphi$ is esb then $\varphi^{sb}$ denotes the formula
$$
\ext x_1\leq t_1\cdots\ext x_k\leq t_k\varphi^{sb}(x_1,\ldots,x_k).
$$
For sequents and inference rules, their sb versions are defined analogously.
\end{definition}

Theorem \ref{thm:wit1} is a corollary to the following theorem.
\begin{theorem}\label{thm:wit2}
  Let $\Gamma\rightarrow\Delta$ be provable in $L^i_2$ and 
$\Gamma^{sb}\rightarrow\Delta^{sb}$ be of the form 
$$
\begin{array}{ll}
\ext x_1\leq s_1A_1^{sb}(\vec{a},x)\land
&\cdots\land\ext x_m\leq s_mA_m^{sb}(\vec{a},x)\\
&\rightarrow
\ext y_1\leq t_1B_1^{sb}(\vec{a},x)\lor\cdots
\lor\ext y_n\leq t_nB_n^{sb}(\vec{a},x)
\end{array}
$$
where $A_1,\ldots,A_m,B_1\ldots,B_n$ are sharply bounded. Then there exist 
functions $f_1,\ldots,f_n\in LD^i$ such that 
$$
\begin{array}{ll}
b_1\leq s_1(\vec{a})&A_1^{sb}(\vec{a},x)\land
\cdots\land b_m\leq s_m(\vec{a})A_m^{sb}(\vec{a},x)\\
&\rightarrow
f_1(\vec{a},\vec{b})\leq t_1(\vec{a})B_1^{sb}(\vec{a},f(\vec{a},\vec{b}))
\lor\cdots\lor
f_n(\vec{a},\vec{b})\leq t_n(\vec{a})B_n^{sb}(\vec{a},\vec{b}),
\end{array}
$$
where $\vec{b}=b_1,\ldots,b_m$.
\end{theorem}
\begin{pf}
  Induction on the number of sequences in the $L^i_2$-proof of the sequent 
$\Gamma\rightarrow\Delta$. The proof is divided into cases according to 
the last inference of the $L^i_2$-proof. It suffices to show that the last 
inference $I$ is $ep\sbi{1}$-$L^{i+1}IND$ since other cases are identical 
to that of Theorem 4.3 in \cite{ct}. Let $I^{sb}$ be of the form 
$$
\begin{array}{c}
\ext x\leq s(a)\varphi(b,a,x)\rightarrow
\ext  y\leq t(a)\psi(a,y),\ext x\leq s(a)\varphi(b+1,a,x)\\
\hline\\
\ext x\leq s(a)\varphi(0,a,x)\rightarrow
\ext  y\leq t(a)\psi(a,y),\ext x\leq s(a)\varphi(|t'|_{i+1},a,x)\\
\end{array}
$$
where $\varphi(b,a,c)$ and $\psi(a,d)$ are sharply bounded in the extended 
language that includes Skolem functions for esb formulae. By the induction 
hypothesis we have $LD^i$ functions $g$ and $h$ which witnesses the upper 
sequent, namely, 
$$
\begin{array}{rl}
c\leq s(a),\varphi(b,a,c)\rightarrow&g(a,b,c)\leq t(a)\land\psi(a,g(a,b,c)),\\
&h(a,b,c)\leq s(a)\land\varphi(b+1,a,h(a,b,c)),
\end{array}
$$
Now define 
$$
G'(a,c)=\mu x\leq|t'|_{i+1}\left(g(a,x,c)\leq t(a)\land\psi(a,g(a,x,c))\right)
$$
and $G(a,c)=g(a,G'(a,c),c)$. Since it is easily seen that the $\mu$-operator 
of the form $\mu x\leq |t|_{i+1}$ can be computed by $LD^i$, $G'$ and hence 
$G$ are in $LD^i$. Define $f$ by $W^{i+1}BRN$ as,
$$
\begin{array}{rl}
F(a,c,0)&=\min(c,s(a))\\
F(a,c,n+1)&=\min(c,h(a,m,F(a,c,n)))\\
f(a,c)&=F(a,c,|t'|_{i+1})
\end{array}
$$
Then $f\in LD^i$ and 
$$
\begin{array}{rl}
c\leq s(a),\varphi(0,a,c)\rightarrow&g'(a,c)\leq t(a)\land\psi(a,g'(a,c)),\\
&f(a,c)\leq s(a)\land\varphi(|t'|_{i+1},a,f(a,c))
\end{array}
$$
holds. So we are done. 
\end{pf}
\begin{pf}[of Theorem \ref{thm:wit1}]
  Let $\varphi$ be esb in $L^i_2$. Let $\Gamma$ be an empty cedent and 
$\Delta\equiv\ext y\leq t\varphi(x,y)$. Then applying Theorem \ref{thm:wit2} 
to the sequent $\Gamma\rightarrow\Delta$ we get the claim.
\end{pf}

\section{A Non-conservation Result for $L^i_2$}
In this section we shall show that for any $i\in\omega$, the theory $L^i_2$ 
is strictly stronger than the weaker theory $AC^0CA$. The proof is based on 
KPT witnessing theory by Kraj\'\i\v cek Pudl\'ak and Takeuti \cite{kpt} 
which utilizes Herbrand's theorem for universal theories.

First we introduce a computation principle that realizes the counter-example 
computation. 

\begin{definition}
  Let $R(x,y)$ be a binary predicate. Then define the predicate 
$$
\begin{array}{rl}
R^{(i)}(x,y,z)&\\
\Leftrightarrow
&|y|_i\leq|x|_i\land(y>0\rightarrow R(x,y))
 \land (|y|_i<|z|_i\leq|x|_i\rightarrow\neg R(x,z)).
\end{array}
$$
The computation principle $\Omega^i(AC^0)$ is the following:

\noindent
For any $R(x,y)\in AC^0$ there exists a finite number of functions 
$f_1,\ldots,f_k\in AC^0$ such that 
$$
\begin{array}{l}
\mbox{either }\all zR^{(i)}(a,f_1(a),z)\mbox{ is true,}\\
\phantom{either zR(i)(a,f_1(a),z)}
\mbox{or if }b_1\mbox{ is such that }\neg R^{(i)}(a,f_1(a),b_1)\\
\mbox{then either }\all zR^{(i)}(a,f_2(a,b_1),z)\mbox{ is true,}\\
\phantom{either zR(i)(a,f_1(a),z)}
\mbox{or if }b_2\mbox{ is such that }\neg R^{(i)}(a,f_2(a,b_1),b_2)\\
\mbox{then }\cdots\\
\mbox{then }\all zR^{(i)}(a,f_k(a,b_1,\ldots,b_{k-1},z))\mbox{ is true.}
\end{array}
$$
\end{definition}

By generalizing the proof in \cite{kpt}, the following holds,
\begin{proposition}
For any $i\in\omega$, $\Omega^i(AC^0)$ implies $NP\subseteq AC^0/poly$. 
\end{proposition}

\begin{pf}
  For simplicity, we will show for the case where $i=3$. The general 
case can be treated similarly. 
Let $A(x)\equiv\all w\leq xB(x,w)$ be an NP-complete predicate via $AC^0$ 
reduction where $B\in AC^0$. Then it suffices to show that there 
exists a function $g\in AC^0$ and a polynomial growth function $h(n)$ 
such that 
$$
A(x)\Leftrightarrow B(x,g(x,h(|x|)))
$$
holds. We say that $w$ witnesses $x$ if $w\leq x\land B(x,w)$ holds. 
Consider the formula 
$$
R(\seq{a_1,\ldots,a_m},\seq{w_1,\ldots,w_n})
\Leftrightarrow m\leq n\land\all i\leq m(\mbox{$w_i$ witnesses $a_i$}).
$$
Then By $\Omega^3(AC^0)$ there exists a finite sequences of functions 
$f_1,\ldots,f_k\in AC^0$ which interactively compute a maximal number of 
witnesses $w_1,\ldots,w_u$ for some initial segment of $a_1,\ldots,a_m$. 

Fix $n\in\Bbb N$ and let $V_1=\{x\ :\ |x|=n\}$. Let $l=2^{2^k}$. 
First we shall construct an algorithm which computes a pair $\seq{j,w}$ 
on input $a=\seq{x_1,\ldots,x_k}\in {V_1}^l$ such that $w$ witnesses $x_j$. 
The algorithm works as follows:

\begin{tabbing}
  {\bf begin}\\
  \quad\= $y=f_1(a)$\\
  \> {\bf if} $len(y)\geq 1$ {\bf and} $R(a,y)$ {\bf then}\\
  \>\quad\= output $\seq{1,(y)_1}$ {\bf and halt}\\
  \> {\bf endif}\\
  \> {\bf for} $i:=2$ to $k$ do\\
  \>\>$y:=f_j(a,b_1,\ldots,b_j-1)$\\
  \>\>{\bf if} $len(y)\geq j$ {\bf and} $R(a,y)$ {\bf then}\\
  \>\>\quad\= $w:=\seq{(y)_{2^{2^{j-2}+1}},\ldots,(y)_{2^{2^{j-1}}}}$\\
  \>\>\> output $\seq{j,w}$ {\bf and halt}\\
  \> {\bf endfor}\\
  {\bf end.}
\end{tabbing}

Let $Q\subset V_1$ be such that $|Q|=l-1$ and $v\in V_1\setminus Q$. 
We say that $Q$ helps $v$ if for some ordering 
$a=\seq{x_1,\ldots,x_{j-1},v,x_{j+1},\ldots,x_{l}}$ of $Q\cup\{v\}$, the 
above algorithm computes a witness $\seq{j,w}$ for $v$. Since $l$ is a 
constant value, it is straightforward to see that there exists an $AC^0$ 
algorithm to check on input $Q$ and $v$ whether $Q$ helps $v$. Furthermore, 
such algorithm can compute a witness $w(Q,v)$ for $v$ if exists. 

Now we shall show that with the help of polynomially many sets 
$Q_1,Q_2,\ldots$ each with size $l-1$, we can compute a witness for 
given $a\in V_1$, thus we can construct an $AC^0$ algorithm which takes 
$a$ and polynomial advice $Q_1,Q_2,\ldots$ as input and compute a witness 
for $a$. 

Let $|V_1|=N$. Note that there are at least $\binom{N}{l}$ pairs of 
$\seq{Q,v}$ such that $Q$ helps $v$ while there are at most $\binom{N}{l-1}$ 
many sets $Q$ of size $l-1$. So there exists a set $Q_1$ which helps at least 
$\frac{N-l+1}{l}$ different elements of $V_1\setminus Q_1$. 
Define the set $V_{i}$ inductively as,
$$
V_{i+1}=\left\{v\in V_i\ :\ Q_i \mbox{ does not help }v\right\}.
$$
Then in general, there exists $Q_{i+1}\subseteq V_{i+1}$ which helps at least 
$\frac{|V_{i+1}|-l+1}{l}$ many elements of $V_{i+1}\setminus Q_{i+1}$. 
Let $t=\min\{i\ :\ |V_i|\leq l\}$. Then since 
$$
\begin{array}{ll}
|V_{i+1}|&=|V_i|-\displaystyle\frac{|V_i|-l+1}{l}
          <\displaystyle\frac{l-1}{l}|V_i|+1\\
         &<\displaystyle\left(\frac{l-1}{l}\right)^i|V_1|+k
\end{array}
$$
it holds that $t=\lceil\log_{l/(l-1)}N\rceil=O(n)$. Let $w(x)$ be an 
canonical witness for $x$ and define $S_i$ by 
$$
\begin{array}{l}
S_i:=\seq{\seq{x,w(x)}\ :\ x\in Q_i} \mbox{ for }i<t,\\
S_t:=\seq{\seq{x,w(x)}\ :\ x\in V_t}.
\end{array}
$$
Finally let $h(n)=S=\seq{S_1,\ldots,S_t}$. Then $|S|=O(ln^2)$. 
Now we can construct an algorithm which computes a witness for $x$ 
with the help of $S$. 
\begin{tabbing}
  {\bf begin}\\
  \quad\= {\bf if} $x$ occurs in $S$ {\bf then} output $w(x)$\\
  \>{\bf else}\\
  \>\quad\={\bf for} $j:=1$ {\bf to} $t-1$ {\bf do in parallel}\\
  \>\>\quad\={\bf if} $Q_j$ helps $x$ then $w_j:=w(Q_j,x)$\\
  \>\>{\bf endfor}\\
  \>\> output $w_j$ with minimal $j<t$\\
  \>{\bf endif}\\
  {\bf end.}
\end{tabbing}
Since this algorithm is in $AC^0$ we are done.
\end{pf}

\begin{corollary}\label{cor:omega}
  $\Omega^i(AC^0)$ is false.
\end{corollary}
\begin{pf}
This follows from the fact that the parity function is not in $AC^0/poly$,  
since H\aa stad's proof \cite{hastad} that parity cannot be computed 
by constant depth polynomial size circuits can be applied to 
nonuniform case.
\end{pf}

\begin{lemma}
  For any $R(x,y)\in AC^0$, $L^i_2\vdash\ext y\all zR^{(i+1)}(a,y,z)$. 
\end{lemma}
\begin{pf}
  Let $R\in AC^0$. Then clearly it is esb in $L^i_2$. So by 
$ep\sbi{1}$-$L^{(i+1)}MAX$ principle in $L^i_2$ there exists a maximal $t$ 
such that 
$$
\ext x\leq a R(a,x)\rightarrow\ext x\leq a(|x|_{i+1}=t\land R(a,x)).
$$
Thus $L^i_2$ proves
$$
\ext x\leq a R(a,x)\rightarrow
\ext x\leq a\all y\leq a R(a,x)
\land(|x|_{i+1}=|y|_{i+1}\rightarrow\neg R(a,y)).
$$
which is logically equivalent to our assertion. 
\end{pf}

Now we conclude that $AC^0CA$ and $L^i_2$ are distinct.
\begin{theorem}
  $L^i_2$ is not a conservative extension of $AC^0CA$. 
\end{theorem}
\begin{pf}
  Let $R(x,y)\in AC^0$ and suppose $AC^0CA=L^i_2$. Then 
$$
AC^0CA\vdash\ext y\all zR^{(i+1)}(a,y,z)
$$ 
and as $AC^0CA$ is an 
universal theory, we may apply Herbrand's theorem to get a finite number 
of functions $f_1,\ldots,f_k$ such that 
$$
R^{(i+1)}(a,f_1(a),b_1)\lor R^{(i+1)}(a,f_2(a,b_1),b_2)\lor\cdots\lor
R^{(i+1)}(a,f_k(a,b_1,\ldots,b_{k-1}),b_k)
$$
holds. Therefore $\Omega^{(i+1)}(AC^0)$ holds and this contradicts to 
Corollary \ref{cor:omega}. Hence $AC^0CA\neq L^i_2$.
\end{pf}

\section{Concluding Remarks}

The system $L^i_2$ has a slightly complicated definition: it has 
$sb\sbi{1}$-$L^{i+1}IND$ as an induction axiom. One might ask what 
if we replace it with simply $\sbi{1}$-$L^{i+1}IND$. I other words, 
what is the computational complexity of $\sbi{1}$-$L^{i+1}IND$. 
Or, more generally, the problem can be stated as follows:
\begin{problem}
  Specify the complexity of $\sbi{k}$-definable functions of 
$AC^0CA+\sbi{k}-L^iIND$.
\end{problem}


\begin{thebibliography}{99}
\bibitem{ajtai}{\sc M.~Ajtai}. {\it Parity and the pigeonhole principle}. 
In S.~R.~Buss and P.~J.~Scott, eds., Feasible Mathematics, Birkh\"auser. 
(1990) 1-24.
\bibitem{buss}{\sc S.~R.~Buss}. {\it Bounded Arithmetic}. Bibliopolis. (1986).
\bibitem{bb}{\sc S.~R.~Buss and M.~L.~Bonet}.
{\it The Serial Transitive Closure Problem for Trees}.
SIAM J. Comput., {\bf 24} (1995) 109-122.
\bibitem{ct}{\sc P.~Clote and G.~Takeuti}. 
{\it First Order Bounded Arithmetic and Small Circuit Complexity classes}. 
In: Feasible Mathematics II, Birkh\"auser. (1995) 154-218 .
\bibitem{fss}
{\sc M. Furst, J.B. Saxe and M. Sipser}.
{\it Parity, Circuits and the Polynomial-Time Hierarchy}. 
Mathematical Systems Theory. (1984). 13--27.
\bibitem{hastad}{\sc J.~H\aa stad}.
{\it Computational Limitations on Small Depth Circuits.} 
Ph.D. thesis, Massachusetts Institute of Technology, 1986.
\bibitem{jp}{\sc J.~Johannsen and C.~Pollet}. 
{\it On $\dbi{1}$ Bit Comprehension Rule}. 
to appear in: Proceedings of Logic Colloquium '98.
\bibitem{krajicek}{\sc J.~Kraj\'\i\v cek}. 
{\it Bounded Arithmetic, Propositional Logic, and Complexity Theory}. 
Cambridge Univ. Press, (1995).
\bibitem{kpt}{\sc J.~Kraj\'i\v cek, P.~Pudl\'ak and G.~Takeuti}.
{\it Bounded arithmetic and the polynomial hierarchy}. 
Ann. Pure and Appl. Logic. {\bf 52} (1998) 143--153.
\bibitem{kuroda1}{\sc S.~Kuroda}. 
{\it On a Theory for $AC^0$ and the Strength of the Induction Scheme}. 
Math. Logic. Quart. {\bf 44} (1998) 417-426. 
\bibitem{kuroda2}{\sc S.~Kuroda}. {\it Function Algebras for Very Small 
Depth Circuits}. submitted.
\end{thebibliography}
\end{document}